\documentstyle[11pt]{article}
\begin{document}
%%%%%%%%%%%%%%%%%%%%%%%%%%%%%%%%%%%%%%%%

\setlength{\leftmargini}{.5\leftmargini}

\newcommand{\ls}[1]
   {\dimen0=\fontdimen6\the\font \lineskip=#1\dimen0
\advance\lineskip.5\fontdimen5\the\font \advance\lineskip-\dimen0
\lineskiplimit=.9\lineskip \baselineskip=\lineskip
\advance\baselineskip\dimen0 \normallineskip\lineskip
\normallineskiplimit\lineskiplimit \normalbaselineskip\baselineskip
\ignorespaces }

%\ls{1} % single space 
%\ls{2} % double space
%\ls{1.6}
%\ls{1.8}

%%% The  Following instruction decreases the length between two items
 %   in the list environments, to be the same as between two paragraphs.
 %\setlength{\itemsep}{0.0in}

 %   The following instruction decreases the length before the first
 %   item in a list to be the same as between two paragraphs.
 %   in the list environments, to be the same as between two paragraph.
 %\setlength{\topsep}{0.0in}

%TO ADD THE FULL-STOP AFTER THE SECTION NO. E.G. 1. INTRODUCTION
\def\thepart{\Roman{part}} 
\def\thesection {\arabic{section}.}
\def\thesubsection {\thesection\arabic{subsection}.}
\def\thesubsubsection {\thesubsection\arabic{subsubsection}.}
\def\theparagraph {\thesubsubsection\arabic{paragraph}.}
\def\thesubparagraph {\theparagraph\arabic{subparagraph}.}

    \setcounter{equation}{0}

\newcommand{\hsp}{{\hspace*{\parindent}}}

\newtheorem{problem}{Problem}[section]
\newtheorem{definition}{Definition}[section]
\newtheorem{lemma}{Lemma}[section]
\newtheorem{proposition}{Proposition}[section]
\newtheorem{corollary}{Corollary}[section]
\newtheorem{example}{Example}[section]
\newtheorem{conjecture}{Conjecture}[section]
\newtheorem{algorithm}{Algorithm}[section]
\newtheorem{theorem}{Theorem}[section]
\newtheorem{exercise}{Exercise}[section]

% put a period after theorem and theorem-like numbers
\makeatletter
\def\@begintheorem#1#2{\it \trivlist \item[\hskip \labelsep{\bf #1\
#2.}]}
\makeatother
                                                   
\newcommand{\be}{\begin{equation}}
\newcommand{\ee}{\end{equation}}
\newcommand{\bea}{\begin{eqnarray}}
\newcommand{\eea}{\end{eqnarray}}

\newcommand{\beq}[1]{\begin{equation}\label{#1}}
\newcommand{\eeq}{\end{equation}}
\newcommand{\req}[1]{(\ref{#1})}
\newcommand{\beqn}[1]{\begin{eqnarray}\label{#1}}
\newcommand{\eeqn}{\end{eqnarray}}

\newcommand{\beaa}{\begin{eqnarray*}}
\newcommand{\eeaa}{\end{eqnarray*}}

\newcommand{\eq}[1]{(\ref{#1})}

\def\le{\leq}
\def\ge{\geq}
\def\lt{<}
\def\gt{>}

\newcommand{\lip}{\langle}
\newcommand{\rip}{\rangle}
\newcommand{\uu}{\underline}
\newcommand{\oo}{\overline}
\newcommand{\La}{\Lambda}
\newcommand{\la}{\lambda}
\newcommand{\eps}{\varepsilon}

\newcommand{\dsum}{\displaystyle\sum}
\newcommand{\dfr}{\displaystyle\frac}
\newcommand{\bige}{\mbox{\Large\it e}}
\newcommand{\integers}{Z\!\!\!Z}
\newcommand{\rationals}{{\rm I\!Q}}
\newcommand{\reals}{{\rm I\!R}}
\newcommand{\realsd}{\reals^d}
\newcommand{\NN}{{\rm I\!\!N}}
\newcommand{\degree}{{\scriptscriptstyle \circ }}
\newcommand{\dfn}{\stackrel{\triangle}{=}}
\def\complex{\mathop{\raise .45ex\hbox{${\bf\scriptstyle{|}}$}
     \kern -0.40em {\rm \textstyle{C}}}\nolimits}
\def\hilbert{\mathop{\raise .21ex\hbox{$\bigcirc$}}\kern -1.005em {\rm\textstyle{H}}} %Hilbert space

\newcommand{\calA}{{\cal A}}
\newcommand{\calC}{{\cal C}}
\newcommand{\calD}{{\cal D}}
\newcommand{\calF}{{\cal F}}
\newcommand{\calL}{{\cal L}}
\newcommand{\calM}{{\cal M}}
\newcommand{\calP}{{\cal P}}
\newcommand{\calX}{{\cal X}}

\newcommand{\Prob}{{\rm Prob\,}}
\newcommand{\mod}{{\rm mod\,}}
\newcommand{\sinc}{{\rm sinc\,}}
\newcommand{\ctg}{{\rm ctg\,}}
\newcommand{\ifff}{\mbox{\ if and only if\ }}
\newcommand{\proof}{\noindent {\bf Proof:\ }}
\newcommand{\remark}{\noindent {\bf Remark:\ }}
\newcommand{\remarks}{\noindent {\bf Remarks:\ }}
\newcommand{\note}{\noindent {\bf Note:\ }}

\newcommand{\boldx}{{\bf x}}
\newcommand{\boldX}{{\bf X}}
\newcommand{\boldy}{{\bf y}}
\newcommand{\uux}{\uu{x}}
\newcommand{\uuY}{\uu{Y}}

\newcommand{\limn}{\lim_{n \rightarrow \infty}}
\newcommand{\limN}{\lim_{N \rightarrow \infty}}
\newcommand{\limr}{\lim_{r \rightarrow \infty}}
\newcommand{\limd}{\lim_{\delta \rightarrow \infty}}
\newcommand{\limM}{\lim_{M \rightarrow \infty}}
\newcommand{\limsupn}{\limsup_{n \rightarrow \infty}}

\newcommand{\imii}{\int_{-\infty}^{\infty}}
\newcommand{\imix}{\int_{-\infty}^x}
\newcommand{\ioi}{\int_o^\infty}

\newcommand{\ARROW}[1]
  {\begin{array}[t]{c}  \longrightarrow \\[-0.2cm] \textstyle{#1} \end{array} }

\newcommand{\AR}
 {\begin{array}[t]{c}
  \longrightarrow \\[-0.3cm]
  \scriptstyle {n\rightarrow \infty}
  \end{array}}

\newcommand{\pile}[2]
  {\left( \begin{array}{c}  {#1}\\[-0.2cm] {#2} \end{array} \right) }

\newcommand{\floor}[1]{\left\lfloor #1 \right\rfloor}

%for doing boldface subscripts etc., e.g. $G_{\mmbox{\boldx}}$
\newcommand{\mmbox}[1]{\mbox{\scriptsize{#1}}}

%fraction with round brackets
\newcommand{\ffrac}[2]
  {\left( \frac{#1}{#2} \right)}

\newcommand{\one}{\frac{1}{n}\:}
\newcommand{\half}{\frac{1}{2}\:}

%qed
\def\squarebox#1{\hbox to #1{\hfill\vbox to #1{\vfill}}}
\newcommand{\qed}{\hspace*{\fill}
           \vbox{\hrule\hbox{\vrule\squarebox{.667em}\vrule}\hrule}\smallskip}

%for the symbol > with a ~ underneath it 

\newcommand{\RAISE}{{\:\raisebox{.7ex}{$\scriptstyle{>}$}\raisebox{-.3ex}
           {$\scriptstyle{\!\!\!\!\!\sim}\:$}}}

\newcommand{\DOWN}{{\:\raisebox{.7ex}{$\scriptstyle{<}$}\raisebox{-.3ex}
           {$\scriptstyle{\!\!\!\!\!\sim}\:$}}}

\title{\bf The role of \boldmath${\nu_{\tau}}$ ultrahigh energy
astrophysics in km$^3$ detectors}

\centerline{
\author{Daniele Fargion\\
Physics Department\\
Rome Univ. 1, I.N.F.N. Rome\ 1\\
Pl. A. Moro\ 2, Italy\\
\\
Faculty of Electrical Engineering\\
Technion---Israel Institute of Technology\\
Haifa \ 32000, Israel}}
\date{Preprint INFN 1166/97\\
Roma I 17/2/97\\
EE.PUP 1074, Feb 1997}

\maketitle

\thispagestyle{empty}
\setcounter{page}{0}
\begin{abstract}
We show that the expected $ \nu_{\tau} $ signals, by their secondary tau tracks,
in $Km^3$ detectors at highest cosmic ray energy window
$ 1.7\cdot 10^{21} \, \mathrm{eV} \gt E_{\tau} \gt 1.6\cdot 10^{17} \, \mathrm{eV}$,
must overcome the corresponding $ \nu_{\mu} $ (or muonic) ones.
Indeed, the Lorentz-boosted tau range length grows (linearly) above muon range,
for $ E_{\tau} \RAISE 1.6 \cdot 10^8 $~GeV and reaches its
maxima extension, $ R_{\tau_{\max}} \simeq$~191~km, at energy
$ E_{\tau} \simeq 3.8\cdot 10^9$~GeV.
At this peak the tau range is nearly 20~times the corresponding muon range
(at the same energy) implying a similar ratio in $ \nu_{\tau} $ over 
$ \nu_{\mu} $ detectability.
This dominance, however may lead (at present most abundant
$ \nu_{\tau} $ model fluxes) to just a rare spectacular event a year (if flavor
mixing occurs).
Lower energetic $ \tau $ and $ \nu_{\tau} $ signals
$ (\bar{\nu}_e e\to \bar{\nu}_{\tau} \tau \, , \; \nu_{\tau} N\to \, \cdots ) $
at energy range ($ 10^5 \div 10^7$~GeV) may be more easily observed in km$^3$
detectors at a rate of a few $ ( \bar{\nu}_e e\to \bar{\nu}_{\tau} \tau) $ to  
tens event $ (\nu_{\tau} N\to \tau + $ anything) a year.
\end{abstract}

\section{Introduction}
\ls{1.5}
High energy astrophysics is waiting for the new neutrino telescope generations
able to reveal the expected TeV (and above) energetic neutrinos ejected by active
nuclei (AGN) blazars \cite{1} as well as from galactic supernova relics or
galactic mini-blazars.
The common theoretical argument in favor for neutrino cosmic ray (c.r.) source
is the last
experimental evidence of extragalactic TeV photon sources (Markarian 421,501) and
the unique neutrino transparency through cosmic 2.75~K$^{\degree}$ B.R.
from cosmic distances.
Secondary atmospheric neutrinos will play a negligible role at high
($\RAISE 10^4\div 10^5$~GeV) c.r. energy.
The common expected neutrinos are of electronic $ ( \nu_e , \, \bar{\nu}_e ) $ and
muonic $ (\nu_{\mu} , \, \bar{\nu_{\mu}} ) $ nature because of the ``low energetic''
pion progenitor masses $ ( \, m_{\pi^{\pm}} ) $, and their consequent
easier and larger productions in proton-proton scattering.
However, at very high energy $ ( E_p \gt 10^{15} \, \mathrm{eV} ) $ the $p$-$p$
scattering may lead, by charm production, to other secondary charmed hadrons able
to decay also in tau leptons; this possibility allows (at least above
$ 10^{15}\, \mathrm{eV}) $ the production of $ \nu_{\tau} $ component as abundant
as $ \nu_e , \, \nu_{\mu} $ ones.
Moreover, flavor mixing and oscillation like
$ (\nu_{\mu} \leftrightarrow \nu_{\tau} ) $,
even at most wide and unexplored parameter ranges
$ (1 \gt \sin^2 2\theta_{\mu\tau}\gt 0 , \, \Delta m_{\mu \; \tau}^2 \ll 0.2 \, \mathrm{eV}^2) $
are well compatible with characteristic large galactic and huge cosmological
lengths $ L_g \sim 10^{24} \, \mathrm{cm} , \, L_c \sim \frac{c}{H_o} \sim 10^{28}
\, \mathrm{cm}$.
Indeed the flavor oscillation length is
\beq{1}
L_{\nu} (\nu_i\to \nu_j) = 1.23\cdot 10^{16} \, \mathrm{cm} \,
\ffrac{E_{\nu}}{10^{20} \, \mathrm{eV}}\, \ffrac{\Delta m_{i, \, j}}{\mathrm{eV}}^{-2} \ll
L_g , \, L_c \; .
\eeq
Therefore flavor mixing may easily lead to an abundant $ \nu_{\tau} $ production.
However, ultrahigh energy $ \nu $ interactions with matter, deeply overviewed and
summarized by last R.~Gandhi, C.~Quigg, M.~H.~Reno, I.~Sarcevic reports \cite{2}
received little attention to the $ \nu_{\tau} $ role (probably because of the very
short unstable lifetime of the $ \tau $ lepton: $ (\tau_{\tau} \sim
\ffrac{m_{\tau}}{m_{\mu}}^5 \, \tau_{\mu} \sim 3\cdot 10^{-13}~\mathrm{sec})$).
Nevertheless, a first important UHE $ \nu_{\tau} $ role at PeV energies has been
noted also recently by J.~Learned and S.~Pakvasa \cite{3}; in particular,
these authors stressed that a characteristic hadronic behaviour
at the initial event
of the $ \nu_{\tau} $ interaction and at the end shower of the lepton $ \tau$
track: a ``double bang'' signal.

Here we underline the dominant and key role of $ \nu_{\tau} , \, \bar{\nu}_{\tau} $
tracks signatures by their secondary tau at much higher energies
$ ( \mathrm{E}_{\nu} \RAISE  10^{17} \div 10^{20}\,\mathrm) eV $ over muon ones because of the large
Lorentz factors and the consequent longer tau tracks.
This relativistic tau ``longevity'' makes the heaviest lepton the most easily
detectable above $ 5\cdot 10^{17} \, \mathrm{eV} $ (or $ 10^{17} \, \mathrm{eV} $
in the rock) in a km$^3$ detector.
Lower energetic $ \nu_{\tau} (10^7 \, \mathrm{GeV} \gt E_{\nu} \gt 10^5 \, \mathrm{GeV}) $ 
may be more easily observed because of a more aboundant primary flux as discussed in the conclusion.
Their discovery may lead to the first ``direct'' evidence for the $ \nu_{\tau} $
existence, may open the most fashionating window at the highest
c.r. astrophysics frontiers and it may prove the deepest secrets of most powerful
cosmic accelerators.

\section{The tau radiation length versus the muon one} 

Muons are commonly known as the most penetrating charged cosmic ray because
their radiation length is much longer (at high energy) than the corresponding
electron one.
Indeed, the muon radiation length at high energy is roughly
$ \ffrac{m_{\mu}}{m_e}^2 $ longer than that of the electron, because
(see Jackson \cite{4}, eq.~15.48) the characteristic leptonic bremsstrahlung
radiation length $ b_L^{-1} $ is found classically:
\beq{2}
b_L^{-1} = \left[\dfr{16}{3}\, Z^2 N \, \ffrac{c^2}{\hbar
c}\,\ffrac{c^2}{m_Lc^2}^2 \, \ln \ffrac{\lambda 192 m_L}{Z^{1/3} m_e}\right]^{-1} \; ,
\eeq
where $N$ is the atomic number density which is proportional to the Avogadro
number times the average density, $ \lambda $ is an a-dimensional factor near unity, 
$ m_L $ is the lepton $ (e,\mu,\tau) $ mass and $Z$ is the target nuclear charge.
Therefore neglecting the ``slow'' logarithmic mass dependence, the radiation
length $ b_L^{-1} $ is mainly proportional to the square of the lepton mass $ m_L $.
The radiation loss by pair production would be, at higher energies, the ruling one
(over bremsstrahlung and over the negligible photo nuclear losses).
Nevertheless, all the radiation lengths grow in
similar form i.e., as the square of the lepton mass $ (\sim m_L^2) $.
The reason of it is in the probability amplitude of the corresponding Feymann
diagram, where an exchange of a virtual photon by a nuclei and by the incoming
relativistic lepton leads to the emission of a high energy photon, or an electron pair.
The process amplitude is roughly proportional (because of the lepton mass presence
in the propagator) to the inverse of the lepton mass $ (m_L) $.
The consequent cross section and its inverse (roughly proportional to the radiation
length) decrease (or grow) consequently as $ \ffrac{m_L}{m_c}^{-2} $ (or
$\ffrac{m_L}{m_c}^2 $) as it has been found classically and experimentally in
Eq.~\req{2}.
Therefore the most penetrating lepton must be the heaviest ones, i.e. the tau
leptons.
On the other hand, the lifetime of the unstable tau lepton, being proportional to
the inverse of the fifth power of its mass $ \tau_c \simeq \ffrac{m_{\mu}}{m_c}^5
\, \tau_{\mu} $, makes its track extremely short: $ c\tau_{\tau} = \gamma_{\tau} 9
\cdot 10^{-3} \, \mathrm{cm} $ (with respect to the muon ones).
At highest energies $ (E_{\tau}\gg 100 \, \mathrm{TeV}) $
the huge Lorentz factor boost
the observed short tau lifetime and increase its value linearly with energy while
the corresponding muon tracks already reached, in the water or in the rock, a
nearly steady maxima (a logarithmic growth) of a few kilometers length.
Consequently, at highest energy
($ E_{\tau} \RAISE 5.6 \cdot 10^8 \, \mathrm{GeV} $ in water,
$ 1.6\cdot 10^8 \, \mathrm{GeV} $ in the rock)
the tau radiation length will be the longest one and the
cosmic tau neutrino rays, $ \nu_{\tau} , \, \bar{\nu}_{\tau} $
(if abundant as other flavors) will be the dominant
source of signals in km$^3$ detectors over other leptons at the same energies.
Finally, as for the muons, also the tau radiation length will reach a maxima
extension at the highest energies
$ ( \sim 4\cdot 10^9 \, \mathrm{GeV} ) $ for two main energy losses:
\newcounter{xxx}
\begin{list}
{\alph{xxx})}{\usecounter{xxx}}
\item
The electromagnetic radiation losses (pair production).
\item
The electroweak interactions and losses with matter (mainly nucleons).
\end{list}
The latter processes is the main restrictive constraint on tau tracks
(in water and rock) at \mbox{$ E_{\tau} \RAISE 5\cdot 10^9~ \, \mathrm{GeV} $}
and it provides a
maxima radiation length comparable to those of the neutrino at same energies
($ \RAISE 200 \mathrm{km} $ in the rock, $ \RAISE 420 \, \mathrm{km} $ in water)
which will be discussed further in detail, below.
The growth of the lepton $ \tau $ radiation length and its (proportional) detectability
leads to a fundamental and dominant role of
$ \nu_{\tau} $ UHE $ (\RAISE 10^8 \, \mathrm{GeV}) $ astrophysics,
in a near future km$^3$ or larger neutrino telescope.
Contrary to present arguments, we remind that the {\em absence\/} of
$ \nu_{\tau} $ fluxes when flavour obscillation are forbidden, 
at lower energies $ ( 10^{11} \div 10^{13} \, \mathrm{eV}) $
has already been considered by us \cite{5} (in absence of flavor oscillations),
in order to bound the properties of any hypothetical heavy fourth neutrino
generation clustered, as cold dark matter, in galactic halos.\\
The large ratio of the $ \tau $ radiation length over those of the muons, reaching
in principle a maximal factor $ \sim \ffrac{m_{\tau}}{m_{\nu}}^2 $
(or at least two order of magnitude), might imply a corresponding ratio in the
detectability of the two leptons at those energies
$ (10^8 \div 10^{10} \, \mathrm{GeV}) $;
however nuclear interactions as shown in more detail in the text make this
 ratio smaller $ (\sim 20 ) $.\\
Finally, the secondary muons ``tail'' $ \mu $ due to $ \tau $ decays 
$(\mu_{\tau} )$ will also
increase by a large fraction
($\sim$ 100\%) the indirect $ \tau $ (and $ \nu_{\tau}$) detectability.
Moreover, the most probable $ \tau $ ($\RAISE$ 60\%) hadronic decay
(and its consequent shower) or its electroweak nuclear shower will lead as it has been
noted \cite{3} to an unambiguous ``hadronic'' jet signature in underground detectors,
contrary to common ``quite'' one-track muon leptonic decays.

\section{The source of high energy tau neutrinos}

As we already noted in the introduction at very high energy
$ (E_p \gt 10^{15} \, \mathrm{eV}) $ $p$-$p$ scattering may lead, by charmed
hadronic production, also to a secondary tau (and a neutrino tau $ \nu_{\tau}$), whose
abundance may be as proliferous as other flavor ones $ ( \nu_e , \, \nu_{\mu}) $.
In most models these neutrinos are expected from Active Galactic Nuclei (AGN) or
blazars \cite{1}.

Ultra-high energy neutrinos $ \nu_e , \, \nu_{\mu} ( \bar{\nu}_e , \bar{\nu}_{\mu}) $
$ ( E_{\nu} \gt 10^{19} \, \mathrm{eV}) $ may also be born copiously by photopion
production of high energy proton (and neutron)
$ ( E_p \gg 10^{19} \, \mathrm{eV} ) $ onto cosmic 2.75K$^{\degree}$ BBR and
galactic radio waves background.
Unfortunately, these abundant photopion productions at $ 10^8\div 10^{12} \,
\mathrm GeV $ cannot in general produce direct tau neutrinos.
Nevertheless, there are other related expected able to lead also in this
interesting energy range $(3\cdot 10^8 \, \mathrm{GeV} \div 10^{12}\,\mathrm{GeV} ) $
to primary or secondary high energy tau and $ \nu_{\tau} $:

\vspace*{-.6cm}
\newcounter{yyy}
\begin{list}
{\alph{yyy})}{\usecounter{yyy}}
\item
Hadronic (charm or beauty) showers due to downward or horizontal high energy
neutrinos $ ( \nu_e , \, \nu_{\mu}) $ interaction (mainly nuclear) in the
Earth by charged neutrino-electron interactions;
$ ( \bar{\nu}_e e \to \bar{\nu}_{\tau}\tau) $
at the resonance $W^-$ mass peak are relevant only at energy peak
$ E_{\nu} \sim 6\cdot 10^6 \, \mathrm{GeV} $ to be discussed at the end.
\item
Flavor oscillations $ \nu_{\mu}\to\nu_{\tau} $ (as well as $ \nu_e\to \nu_{\tau}) $
at the widest and even unexplored parameter ranges:
$ (1\gt \sin^2 2 \theta_{e\tau} \gt 0 ) $,
$ (1\gt \sin^2 2\theta_{\mu \tau} \gt 0 ) $;
$ \Delta m_{i, \, \tau}^2 \ll 0.2 \, \mathrm{eV^2} $ \cite{6}.
For any realistic neutrino mass these parameter rays may be satisfied.
\end{list}
Indeed, flavor oscillations lengths, as already mentioned, even stretched by the
huge Lorentz factor is in general below to characteristic cosmological
$ \frac{c}{H_o} $ distances: (see Eq.~\ref{1}).

Finally we remind that ultrahigh energy tau pairs production, by high energy photon
$ ( E_{\gamma} \RAISE 5\cdot 10^{21} \, \mathrm{eV}) $-photon (B.B.R. at 2.7~K)
Compton Scattering, may also take place, but at a very low rate.

Therefore we shall consider in the following the neutrino and anti neutrino tau
cosmic ray flows as abundant (or comparable) as all the other flavors ones; in
conclusion even in absence of any flavor mixing the $ \nu_{\tau} $ secondary
(or $ \tau $)
must exist by $ \nu_e , \, \nu_{\mu} $ hadronic secondaries more probably
 along an horizontal plane
ring (where the $ \nu_{\tau} N , \, \tau N $ interactions lengths are comparable
to the detector depths).
Astrophysical sources and fluxes for such a high energy
$ ( \RAISE 10^7\div 10^{12} \, \mathrm{GeV}) $ neutrinos have been modeled by
many; we refer mainly to the flux calculated by Stecker and Salamon \cite{7} which
will probably dominate in the energy range 
$ (10^7\div 3\cdot 10^8 \, \mathrm{GeV}) $,
labeled by AGN-SS (in ref\cite{2}, in Fig.~18) due to $p$-$p$ scattering at source; in
this range we {\em must\/} expect primary $ \nu_{\tau} $.
We also refer to the photopion production of cosmic rays and the secondary
neutrino flux $ (\nu_e , \, \nu_{\mu} , \, \bar{\nu}_e , \, \bar{\nu}_{\mu}) $
considered by Yoshida and Teshima \cite{8} either for turn-on time at maximal redshift
$ z=2 $ (labeled by CR--2 in \cite{2}) and redshift $ z=4 $ (labeled by CR--4 in
\cite{2}).
For the last two models the expected neutrino maxima fluxes at the neutrino energy
range $ 3\div 5\cdot 10^9 \, \mathrm{GeV} $ reaches a value of
$ \sim 10^{-18} \, \mathrm{cm}^{-2} s^{-1} sr \div 10^{-17} \, \mathrm{cm}^{-2}
s^{-1} sr^{-1} $ i.e. fluxes comparable to those observed at the same energies in
known cosmic rays on Earth.

\section{Ranges of ultrahigh energy tau lepton}

As we already mentioned the radiation length $ b_{\tau}^{-1} $ for tau lepton, due
mainly to pair production in Eq.~\req{2}, will increase the range of tau tracks
(energy dependent) with respect to corresponding of muons, as soon as the Lorentz boost
$ (\gamma_{\tau} )$ will reach large values $ (\gamma_{\tau} \RAISE 10^8 )$ and as
long as the electro weak interaction with nucleons will not bind their growth.

The radiation length $ b_{\tau}^{-1} $ will play a role in defining the tau range
by the general energy loss equation:
\beq{3}
-\dfr{dE_{\tau}}{dx} = a(E_{\tau}) + b_{\tau} (E_{\tau}) E_{\tau} \; ,
\eeq
where $a$ and $b$ are slow energy variable functions respectively for ionization
and radiation losses.
The asymptotic radiation length $ b_{\tau}^{-1} $ at high energies
$ E_{\tau} \gg 10^{15} \, \mathrm{eV} $ is related to the corresponding muon one
by this approximated relation derived by classical bremsstrahlung formula in
Eq.~\req{2} scaled for the two different lepton masses:
\beq{4}
b_{\tau} \simeq \ffrac{m_{\mu}}{m_{\tau}}^2 \cdot \left[\dfr{\ln\,\left(\dfr{\la 192\,
m_{\tau}}{z^{1/3} m_e}\right)}{\ln\left(\dfr{\la 192\, m_{\mu}}{z^{1/3} m_e}\right)}
\right]\cdot b_{\mu} \cong \dfr{b_{\mu}}{219} = 1.78
\cdot 10^{-8} \, \mathrm{cm}^{-1} \rho_r^{-1} \; ,
\eeq
where, $ \rho_r^{-1}$ stand for relative adimensional  density in water unity, 
and where
in the present energy range, $ E_{\tau} \gg 10^5 \, \mathrm{GeV} $, we
assumed that the experimental phenomenological coefficient as in Ref.~\cite{2}:
$ b_{\mu} \simeq 3.9 \cdot 10^{-6} \, \mathrm{cm}^{-1} \rho_r^{-1} $.
The corresponding radiation length $ b_{\tau}^{-1} $ is:
$ b_{\tau}^{-1} \simeq \dfr{561\, \mathrm{km}}{\rho_r} \; , $
%where $ \rho_r $ is the adimensional target density (in water unity).
The ionization coefficients values are: $ a_{\tau} \simeq a_{\mu} = 2 \cdot 
10^{-3} \mathrm{GeV cm} \rho_r^{-1} $.
The integral of the energy loss equation will lead, from the radiation length
$ b_{\tau}^{-1} $, to a larger, energy dependent, radiation range $ R_{R_{\tau}} $:
\beq{5}
R_{R_{\tau}} \equiv\dfr{b_{\tau}^{-1}}{\rho_r} \, \ln\,\dfr{a_{\tau} + b_{\tau}
E_{\tau}}{a_{\tau} + b_{\tau} E_{\tau}^{\min}} \simeq
\dfr{b_{\tau}^{-1}}{\rho_r} \ln \, \dfr{E_{\tau}}{E_{\tau}^{\min}} \; .
\eeq
The last approximation occurs because of the smallness
(for $ E_{\tau} \gg 10^5 \, \mathrm{GeV}$)
of the ionization factor $ a_{\tau} $ with respect to
$ b_{\tau} E_{\tau}^{\min} $ and
$ b_{\tau} E_{\tau} $ terms.

In the Earth, according to the preliminary Earth Model \cite{1} on the first few km
the relative density $ \rho_r $ is unity in the sea, near 3 in the early depth rocks,
around 5 in the first 1000~km Earth depths.
Therefore the consequent tau radiation length from Eqs.~\req{4}--\req{5}
(for $ \rho_r \sim 5), E_{\tau} \gg 10^4 \mathrm{GeV} $ becomes: 
\beq{6}
R_{R_{\tau}} \cong 1292 \, \mathrm{km} \, \left(\dfr{\rho_r}{5}\right)^{-1} \,
\left\{ \dfr{\left[\ln\left(\dfr{E_{\tau}}{10^8 \,
\mathrm{GeV}}\right)\left(\dfr{E_{\tau}^{\min}}{10^4 \,
\mathrm{GeV}}\right)^{-1}\right]}{(\ln \, 10^4 )}\right\} \; .
\eeq
This extreme propagation range, comparable even to the Earth radius, is to be
combined with and bounded by, the more restrictive tau lengths due to short tau
lifetime, as well as by the range due to electro weak tau-nucleons interactions at
the highest energies $ (E_{\tau} \RAISE 10^9 \, \mathrm{GeV} )$.

The role of tau lifetime and its free path length $ R_{\tau_o} $, boosted by
large Lorentz factors $ \gamma_{\tau} =\frac{E_{\tau}}{m_{\tau} c^2} $, grows
linearly with energies:
\beq{7}
R_{\tau_o} = c \tau_{\tau} \gamma_{\tau} = 5 \, \mathrm{km} \,
\left(\frac{E_{\tau}}{10^8 \, \mathrm{GeV}}\right) \; .
\eeq
The electroweak tau-nucleon interaction range, $ R_{W\tau} $, on the other side,
decreases with tau energies in analogy with the corresponding ones for
neutrino-nucleon scattering.
In a first approximation the cross sections
$ \sigma (\nu_{\tau} N) $, at energy of interest
$ 10^6 \, \mathrm{GeV} \le E_{\nu_{\tau}}\le 10^{12} \, \mathrm{GeV} $
may be described by a simple power law form, either for charged and neutral
currents; because of the crossing symmetry in the Feynman diagrams we may also
write (following \cite{2}) similar expressions for $ \sigma (\tau N) $:
\beq{8}
\begin{array}{lll}
\sigma_{cc} (\tau N) & \simeq & \sigma_{cc} (\nu_{\tau} N) = 4.44 \cdot 10^{-33}
\, \mathrm{cm}^2 \, \left(\dfr{E_{\tau}}{10^8 \, \mathrm{GeV}}\right)^{0.402}\\[.4cm]
\sigma_{Nc} (\tau N) & \simeq & \sigma_{Nc} (\nu_{\tau} N) = 1.95 \cdot 10^{-33} \,
\mathrm{cm}^2 \,\left(\dfr{E_{\tau}}{10^8 \, \mathrm{GeV}}\right)^{0.408}\\[.4cm]
\sigma_{cc} (\bar{\tau} N) & \simeq & \sigma_{cc} (\bar{\nu}_{\tau} N) =
4.3 \cdot 10^{-33} \, \mathrm{cm}^2 \, \left(\dfr{E_{\tau}}{10^8 \,
\mathrm{GeV}}\right)^{0.404}\\[.4cm]
\sigma_{Nc} (\bar{\tau} N) & \simeq & \sigma_{Nc} (\bar{\nu}_{\tau} N) =
1.87 \cdot 10^{-33} \, \mathrm{cm}^2 \,\left(\dfr{E_{\tau}}{10^8 \,
\mathrm{GeV}}\right)^{0.41}
\end{array}
\eeq
The corresponding averaged electroweak range $ R_{W\tau} $ in the energy range of
interest in water for a total (charged + neutral) cross sections
$ \sigma (\tau N ) \simeq \sigma (\bar{\tau} N) \simeq 6.5 \cdot 10^{-33} \,
\mathrm{cm}^2 \, \ffrac{E_{\tau}}{10^8 \, \mathrm{GeV}}^{0.404} $ is:
\beq{9}
R_{W_{\tau}} = \dfr{1}{\sigma N_A \rho_r} \simeq \dfr{2.5\cdot 10^3 \,
\mathrm{km}}{\rho_r} \,
\left(\dfr{E_{\tau}}{10^8\, \mathrm{GeV}}\right)^{-.404} \; .
\eeq
The total tau range, $ R_{\tau} $, is just the minimal value of the three above ones:
the radiation one $ R_{R_{\tau}} $ in Eq.~\req{5}, the lifetime one
$ R_{\tau_o} $ in Eq.~\req{7}, the present electroweak-nuclear one $ R_{W\tau} $
in Eq.~\req{9}:
\beq{10}
R_{\tau} = \left( \dfr{1}{R_{R_{\tau}}} +
\dfr{1}{R_{\tau_o}} +\dfr{1}{R_{W\tau}}\right)^{-1} \; .
\eeq
Let us notice that in the estimate of the electroweak range $ R_{W\tau} $ we
neglected the (otherwise) interesting electron-tau electroweak interactions in
the atoms for two main reasons:
\newcounter{zzz}
\begin{list}
{\alph{zzz})}{\usecounter{zzz}}
\item
The tau-electron electroweak cross sections $ (\tau e\to \tau e) $ do not 
experience the (corresponding) resonant peak (as for neutrino-electron
scattering: $ \nu_{\tau} e\to W^-\to \tau {\nu_e} $) at energies
$ E_{\nu} \sim 6.10^{15} \, \mathrm{eV} $.
The analogous resonant reaction $ (\tau^+ e\to Z_o\to \nu_{\tau}
 \nu_e) $ is forbidden by flavor conservation number.
Tau and electrons may only interact weakly by electroweak exchange of a
neutral virtual boson $ Z_o$ and a photon.
\item
Even in the above (not allowed) case of a resonant cross section at energy
$ E_{\nu} \sim 6\cdot 10^{15} \, \mathrm{eV} $ and cross section $ \sigma_{\tau
e} \sim 10^{-31} \, \mathrm{cm}^2 $, the shortest tau lifetime and its range
$ R_{\tau_o} $, in Eq.~\req{10}, will mask and hide the short range
$ R_{W\tau} $ (due to hypothetical $ e\tau $ ``resonant'' scattering).
\end{list}
It is important to consider the tau range $ R_{\tau} $ at its characteristic
regimes: when its value will overcome the corresponding muon one
$ R_{\mu} \; (R_{\mu} = R_{\tau}) $, when it will reach its maximal extension
$ R_{\tau} = R_{\tau\max} $, when it will be confined (because of nuclear
interactions) at the highest energies to the same ranges as muon tracks
$ (R_{\tau} = R_{{W}_{\tau}} = R_{\mu}) $.

\section{The critical energies for \boldmath ${\nu_{\tau}}$ dominance}

Let us define the first critical energy, $ E_{\tau_1} $, where the tau range
equals the muon one: $ R_{\tau} = R_{\mu} $ from (Eq.~\req{10}), 
(Eq.~\req{5}) and (Eq.~\req{2}) by substitution of $ b_{\tau} $ with $ b_{\mu}$.
This equation may be easily solved noticing that at this energy range
$ (E_{\tau}\sim 10^8 \, \mathrm{GeV}) $ the shortest and main tau range is the
 lifetime one,
$ R_{\tau} \simeq R_{\tau_o} $; therefore the equation $ R_{\tau} = R_{\mu} $
can be written as follows:
\beq{11}
R_{\tau} \simeq R_{\tau_o} =
c \tau_{\tau} \left(\frac{E_{\tau}}{m_{\tau}{c^2}}\right) =
\dfr{1}{b_{\mu} \rho_r} \, \ln \, \left(\dfr{a+b_{\mu} E_{\mu}}{a+b_{\mu}
E_{\mu}^{\min}}\right) \; ;
\eeq
where one imposes $ E_{\mu} = E_{\tau} $;
numerically one finds that
\beq{12}
5 \, \mathrm{km} \; \left(\dfr{E_{\tau}}{10^8 \, \mathrm{GeV}}\right) \cong
\dfr{2.56 \,}{\rho_r} \, \ln \left(\dfr{E_{\mu}}{E_{\mu}^{\min}}\right) \,
\mathrm{km} \; .
\ee
defines the critical energy $ E_{{\tau}_1} $ where tau track exceedes the 
muonic one.
 For water $ \rho_r = 1 $, and rock
(in these depths $ \langle\rho_r\rangle = 3 $)
the critical energy $ E_{\tau_1} $ and the tau range
$ R_{\tau_1} $ are:
\beq{13}
\begin{array}{llll}
E_{{\tau}_1} = 5.6 \cdot 10^8 \, \mathrm{GeV}\hspace*{.5cm} &&
E_{\tau_1} \cong 1.65 \cdot 10^8 \, \mathrm{GeV} & \\
&\mathrm{(water);}\hspace*{.5cm} && \hspace*{.5cm}\mathrm{(rock)}\\
R_{{\tau}_1} = 28 \, \mathrm{km} &&R_{\tau_1} \cong 8.2\, \mathrm{km}\hspace*{.5cm} &\\
\end{array}
\quad .
\eeq
Here we considered $ E_{\mu}^{\min} = 10^4 \, \mathrm{GeV} $ as in \cite{1}. 
Let us remind that the analytical curve we are using in Eq.~\ref{5} for muons is
a bit overestimated with respect to a more detailed study 
(Lipari, Stanev \cite{9})
and therefore the present critical ``analytical'' value $ E_{\tau_1} $ and the range
$ R_{\tau_1} $, might be larger than the real one (by a factor $ 1.5 \div 2$). 
Therefore from energies $ E_{\tau} \gt 10^8 \, \mathrm{GeV} $ above the tau signal will
overcome the muon ones.
Moreover, the prompt secondary muons from tau decays or from tau hadronic pions
decays (let us label them $ \mu_{\tau} $), may in principle ``double'' the 
expected muonic fluxes;
finally the characteristic tau hadronic  decay (``bang'' \cite{3})  may leave a unique signal.

The linear growth of the tau range $ R_{\tau} $, {\em in absence\/} of the
$ \tau N $ interactions, would reach a maximal radiation range $ R_{T_{\tau}} $
(due to maximal $ b_{\tau}^{-1} $ in Eq.~\req{4}) as in Eq.~\req{6},
at least
two orders of magnitude larger than the corresponding muon range $ ( R_{\mu} ) $.
Indeed, it is possible to show that in such an {\em ideal\/}
(no-electroweak interactions) case the relation $ R_{\tau_o} = R_{\tau_R} $ would
define an extreme energy $ E_{\tau} \simeq 4\cdot 10^{10} \, \mathrm{GeV} $ and a
corresponding range $ R_{\tau_o} \simeq$2000~km, much longer than the $ R_{\mu R} $
range (at the same energy in the rock):
$ R_{\mu} (4\cdot 10^{10} \, \mathrm{GeV})\cong 14 \, \mathrm{km} $.

However, the {\em real\/} maximal tau range is bounded by the the more
restrictive electroweak cross sections (in Eq.~\req{8})
(as for neutrinos) and its range $ R_{W\tau} $ (in Eq.~\req{9}).
The maximal tau range $ R_{\tau_{\max}} $ is then defined by equal conditions in
Eqs.~\req{7} and \req{9}, $ (R_{\tau_o} = R_{W\tau}) $:
\beq{14}
R_{\tau_o} = 5 \, \mathrm{km} \; \left(\frac{E_{\tau}}{10^8 \,
\mathrm{GeV}}\right) = \dfr{2.5\cdot 10^3 \, \mathrm{km}}{\rho_r} \,
\left(\dfr{E_{\tau}}{10^8 \, \mathrm{GeV}}\right)^{-0.404} \; ,
\eeq
whose solution is (for $ \rho_r = 3 $ as in the few hundred terrestrial km depths):
\beq{15}
R_{\tau_{\max}} = 191 \, \mathrm{km} \,
\left(\dfr{\rho_r}{3}\right)^{\frac{1}{1.404}} \, ; \; E_{\tau_{\max}} = 3.8 \cdot 10^9 \, \mathrm{GeV} \;
\left(\dfr{\rho_r}{3}\right)^{\frac{1}{1.404}} \; .
\eeq
For the peculiar case (``horizontal'' neutrinos arrivals), in the sea,
where we may assume
$ \rho_r = 1 $, at a detector depth $ \sim 10 \, \mathrm{km} $,
and a sea $\sim$ 20~km depth, one gets
$ R_{\tau_{\max}} \cong 418 \, \mathrm{km} \, , \; E_{\tau_{\max}} \cong
8.36 \cdot 10^9 \, \mathrm{GeV} $.

It is clear that from Eq.~\req{15} the total maximal tau range $ R_{\tau_{\max}} $
extends 20 times the corresponding muon range at the same energies.
%The length ratio $ \ffrac{R_{\tau_{\max}}}{R_{\mu}} \cong 20 $
%is therefore smaller than the {\em ideal\/} case
%$ \ffrac{R_{\tau R}}{R_{\mu}} \RAISE 150 $, but still quite large.

Finally at higer energies also the electroweak interaction
will bound the muon radiation range and it will make comparable both the taus and muons
ranges.
This will occur once the relation $ R_{\tau} \simeq R_{W\tau} = R_{\mu} $ at the
same energy $ E_{\mu} = E_{\tau} $ (in Eqs.~\req{2}--\req{5} and Eq.~\req{9}) is
satisfied; i.e. when
\beq{16}
R_{\mu} \cong \dfr{b_{\mu}^{-1}}{\rho_r} \, \ln\, \dfr{E_{\mu}}{E_{\mu}^{\min}} =
R_{W\tau} = \dfr{1}{\sigma_W N_A \rho_r} \; .
\eeq
This relation implies, for an adimensional density $\rho_r = 3 $ a critical
energy and ranges:
\beq{17}
E_{\tau_2} \simeq 1.7\cdot 10^{12} \, \mathrm{GeV} \, ; \;
R_{\tau_2} \simeq R_{W\tau} = R_{\mu} = 16 \, \mathrm{km} \; .
\eeq
Therefore from energy $ E_{\tau_1} = 1.6\cdot 10^8 \, \mathrm{GeV} $ up to the
energy
$ E_{\tau_2} \simeq 1.7\cdot 10^{12} \, \mathrm{GeV} $, the tau tracks will
overcome the muon lengths and it will imply a dominant role for tau neutrino
astrophysics (assuming, of course, a  correponding $ \nu_e , \, \nu_{\mu} $ spectra).
Will this dominance be detectable in km$^3$ detectors?
The answer, of course, depends on the unknown primordial cosmic flux: assuming,
as in \cite{2} a model flux labeled CR--2 and CR--4, whose event rates are
summarized in Table~6 \cite{2}, the consequent tau event rate may be promptely 
derived by scaling
the $ R_{\tau} $ range in place of $ R_{\mu} $ for effective km$^3$ volume 
made by the effective area A:
$ A\langle R\rangle $. The most optimistic rates for downward neutrinos
($D$-parton distribution, a CR--4 model \cite{2})
above $ 10^7 \, \mathrm{GeV} $ may reach a muon event rate a year of $ \sim 
4.8 \cdot
10^{-2} $ and a corresponding tau rate just near the unity  for the longest
tau tracks at its maximal range extension
$ R_{\tau_{\max}} $ in Eq.~\req{15}.
Therefore the $ 10^8\div 10^{12} \, \mathrm{GeV} $ energy window dominance of
$ \nu_{\tau} $ is at present models just at the edge of detectability at near future km$^3$
detectors.
However, more yet unobserved and abundant ultrahigh neutrinos fluxes, may 
increase drastically our
predictions.
The $ \nu_{\tau} $ and tau presence in the km$^3$ detectors may
also be discovered by other indirect effects: for instance the presence of 
secondary relic
muon bundles; indeed the hadronic jets due to $ \nu_{\tau} $ nucleon interactions
may also lead to secondary taus, pions and muons whose last tracks (
contemporary  and parallel muons at
a near distance of a few tens of meters)
may prove their common tau decay origin.

\section{The resonant \boldmath${\bar{\nu}_e e\to \bar{\nu}_{\tau} \tau} $ and
\boldmath${\bar{\nu}_{\tau} e\to \tau\nu_e} $ events}

Finally it is worthful to mention that the important and detectable tau
contribution in downward $ \bar{\nu}_e e\to W^-\to \bar{\nu}_{\tau} \tau $ events at
energy tuned at the resonant $ W^- $ formation mass in $ \bar{\nu}_e e$
collisions:
$ E_{\nu}^{\mathrm{res}} = \frac{M_W^2}{2m_e} = 6.3\cdot 10^{15} \, \mathrm{eV}
$; at these ranges of energies the muon range is a few kms long while the tau
range $ R_{\tau} \sim R_{\tau_o} $ due to an average secondary energy
$ \langle E_{\tau}\rangle \simeq \frac{1}{4}\, E_{\nu} = 1.4\cdot 10^{15} \,
\mathrm{eV} $ is only $ R_{\tau} \simeq 71 \, \mathrm{m} $.
Therefore tau ranges are nearly two orders of magnitude smaller than the those of
the muons.

The expected downward muon number of events $ N_{eV} (\bar{\nu}_e
e\to\bar{\nu}_{\mu} \mu ) $ in this resonant energy range, in
0.2~km$^3$ detectors, (see Table~7, \cite{2}) was found to be
$ N_{eV} = 4\div 7 $ a year.
We expect a comparable number of reactions
$ (\bar{\nu}_e e\to \bar{\nu}_{\tau} \tau) $;
however only those events whose originations are confined in 0.2~km$^3$ volume 
will be easely
recognized as tau ones.
Their probability  is reduced by a factor related to the corresponding 
probability to see a
confined $ \mu$ track inside a km size
$ \sim\frac{\mathrm{km}}{L_{\mu}} \sim\frac{1}{5} $;
therefore roughly an event a year due to reaction
$ \nu_e e\to \bar{\nu}_{\tau} \tau $
{\em might be\/} noticed by its tau precursor hadronic shower
(first ``bang'', see \cite{3}) and by its probable (64\%) secondary contained
hadronic cascade (second ``bang'') as well as by its characteristic range (70m).
This (rare) event may occur even in absence of any $ \nu_{\tau} $ cosmic ray and
{\em any\/} neutrino flavor oscillation.
It will be most probable on horizontal tracks where depth size $ \simeq$
interaction length; it will not inform us on any important
$ \nu_{\tau} $ astrophysics nature, but it must prove, at minimally theoretical
assumptions, the same reality of $ \nu_{\tau} $ existence.

Moreover assuming $ \nu_{\tau} $ ultrahigh energy cosmic primary rays at energy range
$ 10^7 \gt E_{\nu_{\tau}} \gt 10^4 \, \mathrm{GeV} $,
the same nuclear electroweak $ \nu_{\tau} N $ interactions, (which lead to
$ \nu_{\tau} $ (and $\tau$) opacity through the Earth at highest energies), are
a source of tau secondaries (even in the energy range where the tau tracks are not
longer than the muon ones).
Indeed, in the energy range $ 10^5 \DOWN \ffrac{E_{\nu}}{\mathrm{GeV}}\DOWN 10^7 $ the
tau production (by $ \nu_e N\to \tau +$ anything) is almost identical to the muon one.
The only difference is due to the range length of tau
$ R_{\tau} \simeq 50 \, \mathrm{m} \, \ffrac{E_{\tau}}{10^6 \, \mathrm{GeV}} $ to
be compared with a few kilometer of a muon $ \mu $ radiation range (very sensitive to
the exact $ E_{\mu}^{\min} $ cut-off).

As before the (detector size/muon range) ratio will offer a first estimate of the
ratio of tau/muon contained signals:
$ R\simeq \frac{1}{5}=$~20\%.
(The $ \nu_e N $ event rate is {\em not\/} suppressed by a much lower ratio
$ \frac{R_{\tau}}{R_{\mu}} \simeq$~1\%).
Therefore nearly 20\% of the corresponding $ \nu_{\mu} , \, \bar{\nu}_{\mu} $
events, expected in a km$^3$ telescope, may be associated to tau signals.
Only 18\% (of these 20\% of events) will mask their tau nature by a
$ \tau\to\mu\nu_{\tau} \nu_e $ decay, nearly at the same energy direction and
therefore hidden in a unique muonic track.
Most (82\%) of the above events will mark their identity by a 50m
$ \ffrac{E_{\tau}}{10^{16} \, \mathrm{GeV}} $ tau precursor track either with a
spectacular and characteristic tau-hadronic shower (a jet) ($\sim$64\%) or by a
short and intense electron shower (whose length, by Landau, Pomeranchuck -- Migdal
effect \cite{9}, is  as short as $ R_e \cong 4m \ffrac{E_{\nu}}{6\cdot 10^{15} \,
\mathrm{eV}}^{1/2}$) or, as noted in \cite{3} by their double ``bangs''.
From the arguments above we nearly expect $\sim 20$ atmospheric events a year to
be associated with $ \tau$ precursor tracks.
Moreover, other $ \sim 20 $ tau events may bring the imprint (and direction)
of primary $ \nu_{\tau} $ cosmic rays born in Active Galactic Nuclear (AGN) or
mini-galactic jets.
These expectations may reach hundred events a year for most optimistic and
proliferous spectra of $ \nu_{\tau} $ sources (see \cite{2}).
Let us remind that here we neglected all other additional hadronic secondary
by ( $ \nu_e , \, \nu_{\mu} $ nucleon elettroweak interactions ) 
showers that may also decays (by charm or
beauty states) in tau leptons.

\section*{Conclusions}

A few tau signals a year in a km$^3$ detector {\em must\/} occur:
\newcounter{qqq}
\begin{list}
{\alph{qqq})}{\usecounter{qqq}}
\item
at $ E_{\nu_{\tau}} \sim 6\cdot 10^{15} \, \mathrm{eV} $ energy range, due to
$ \bar{\nu}_e e\to \bar{\nu}_{\tau} \tau $ resonant event,
disregarding any primary neutrino $
\nu_{\tau} $ source or even in absence of flavor oscillations.
\item
at ``low'' energies $ (E_{\nu} \simeq 10^5 \div 10^6 \, \mathrm{GeV}) $ for any
very probable \cite{1} $ \nu_{\tau} $ primary cosmic rays as abundant as 
$ \nu_e , \, \nu_{\mu} $ ones 
($\nu_{\tau}$ due to charmed hadronic interactions in the source or due to
$ \nu_e \leftrightarrow \nu_{\tau} , \, \nu_{\mu}\leftrightarrow \nu_{\tau} $
flavor oscillations); we expect from ref \cite{2} and the above approximated
arguments, tens of such a
$\nu_{\tau} $ event a year in km$^3$ detectors.
\item
At highest energies, a very rare tau signal a year may probe the dominant 
tau range $ 10^{12} \, \mathrm{GeV} \gt E_{\tau} \gt 10^8 \, \mathrm{GeV} $.
In general, it will cross from size to size the km$^3$ detector, but a few
huge hadronic shower in the km$^3$ detectors (comparable to those observed in
the  rarest atmospheric events $ (\sim 3\cdot 10^{20} \, \mathrm{eV})$),
may leave a unique imprint: a huge Cerenkov flash (at peak power of Megawatt) due to
an initial hadronic shower followed by a collinear (tau) track, whose extension
may easily escape the same km$^3$ detector size.
However, this rare primary ultrahigh $ (E_{\nu} \gg 10^8 \, \mathrm{GeV} $) neutrino event might
be ruled by photopion relics $ (\nu_{\mu} , \, \nu_e) $ 
and therefore it calls for an efficient neutrino flavor oscillation even at widest
allowable parameter ranges (see Eq.~\req{1}, and \cite{3}) during the neutrino 
propagation in the Universe.
\end{list}
In conclusion, we believe that in future km$^3$ telescope more surprises
may (and must) come from neutrino tau and tau signals: the first direct $ \nu_{\tau} $
experimental evidence,  its possible flavor mixing and the first possible spectacular
insight at highest energetic $ (\RAISE 10^8\div 10^{11} \, \mathrm{GeV})$
neutrino astrophysical frontiers.

\paragraph{Acknowledgements}
I would like to thank Prof. M. Lusignoli, Prof. A. Dolgov of Max Born Institute,
Drs. B.~Mele and Dr. A.~Salis, Dr. R. Conversano, Dr. A. Aiello for discussions; in particular I wish to thank Prof. 
A. Dar for and Prof. S. M. Bilenky
important encouragement and suggestions and the Faculty of Electrical Engineering
at Technion for warm hospitality.

\end{document}